\documentclass[conference,onethmnum]{IEEEtran}
\usepackage[utf8]{inputenc} 
\usepackage[T1]{fontenc}
\usepackage{url}
\usepackage{subfig}
\usepackage{ifthen}
\usepackage{cite}
\usepackage[cmex10]{amsmath} % Use the [cmex10] option to ensure complicance
\usepackage{booktabs}
\usepackage{bbm,bm}
\usepackage{enumerate}
\usepackage{enumitem}
\usepackage{graphicx} % Required for inserting images
\usepackage{amssymb}
\usepackage{amsthm}
\usepackage{xcolor}
\usepackage{todonotes}
\usepackage{cleveref}
\usepackage{subcaption}
\theoremstyle{plain}
\newtheorem{theorem}{Theorem}

\newtheorem{definition}[theorem]{Definition}

\crefname{corollary}{Corollary}{Corollaries}

\newcommand{\rank}{\mathrm{rank}}

\let\oldsqrt\sqrt % rename \sqrt as \oldsqrt
\def\sqrt{\mathpalette\DHLhksqrt} % define the new \sqrt in terms of the old one
\def\DHLhksqrt#1#2{%
\setbox0=\hbox{$#1\oldsqrt{#2\,}$}\dimen0=\ht0
\advance\dimen0-0.2\ht0
\setbox2=\hbox{\vrule height\ht0 depth -\dimen0}%
{\box0\lower0.4pt\box2}}

\begin{document}
\title{Robust Spectral Recovery  for   Dynamical Sampling} 
% Robust Spectral Identification in Convolutional Dynamical Sampling with Time-Sparse Corruptions

% Robust Spectrum Recovery for Convolutional Dynamical Sampling under Intermittent Snapshot Corruptions

% Hankel-Robust Spectrum Recovery for Convolutional Dynamical Sampling with Outlier Snapshots

%"Property Inheritance in Tensor Train Subtensors: Incoherence, Condition Numbers, and Rank Preservation"
%"Tensor Train Subtensors: A Study of Property Inheritance and Computational Efficiency"
%"Incoherence and Condition Number Inheritance in Tensor Train Dimensionality Reduction"
% %%% Single author, or several authors with the same affiliation:
% \author{%
%  \IEEEauthorblockN{Author 1 and Author 2}
% \IEEEauthorblockA{Department of Statistics and Data Science\\
%                    University 1\\
 %                   City 1\\
  %                  Email: author1@university1.edu}% }

%%% Several authors with up to three affiliations:
      
\author{%
  \IEEEauthorblockN{HanQin Cai}
  \IEEEauthorblockA{School of Data, Mathematical, and Statistical Sciences  \\ Department of Computer Science \\
                    University of Central Florida\\
                    Orlando, FL 32816, USA\\
                    Email: {hqcai@ucf.edu}%\href{mailto:hqcai@ucf.edu}{hqcai@ucf.edu}
                    }
  \and
      \IEEEauthorblockN{~~~~}
   \IEEEauthorblockA{~~~~~}
  \and \IEEEauthorblockN{Longxiu Huang}
  \IEEEauthorblockA{Department of  %CMSE
  Computational Math, Science and Engineering 
  \\  Department of Mathematics\\
                   Michigan State University\\
                   East Lansing, MI 48824, USA\\
                    Email: {huangl3@msu.edu}%\href{mailto:huangl3@msu.edu}{huangl3@msu.edu}
                    }
   \and 
  \IEEEauthorblockN{Tianming Wang}
  \IEEEauthorblockA{ School of Economic Mathematics\\
  Southwestern University of Finance and Economics\\
Chengdu, Sichuan, China\\
                    Email: {wangtm@swufe.edu.cn}%\href{mailto:hqcai@ucf.edu}{hqcai@ucf.edu}
                    }
   \and
   \IEEEauthorblockN{~~~~}
   \IEEEauthorblockA{~~~~~}
  %  \and \IEEEauthorblockN{~~~~}
  % \IEEEauthorblockA{~~~~~}
  % \and  \IEEEauthorblockN{~~~~}
  % \IEEEauthorblockA{~~~~~}
   \and  
  \IEEEauthorblockN{Juntao You}
  \IEEEauthorblockA{
  School of Artificial Intelligence\\
  National Center for Applied Mathematics in Hubei\\  
  %Hubei Key Laboratory of Computational Science\\
                   Wuhan University\\
                   Wuhan, Hubei, China\\
                    Email: {youjuntao@whu.edu.cn}%\href{mailto:youjuntao@whu.edu.cn}{youjuntao@whu.edu.cn}
                    }
    % \and
  % \IEEEauthorblockN{~~~~}
  % \IEEEauthorblockA{~~~~~}
}

\maketitle

%%%%%%
%% Abstract: 
%% If your paper is eligible for the student paper award, please add
%% the comment "THIS PAPER IS ELIGIBLE FOR THE STUDENT PAPER
%% AWARD." as a first line in the abstract. 
%% For the final version of the accepted paper, please do not forget
%% to remove this comment!
%%
 
 \begin{abstract}
 We study the spectral recovery problem for dynamical sampling on a finite cyclic grid. Given time snapshots obtained from a fixed uniform spatial subsampling of the orbit $x_{\ell}=A^{\ell}f$, we aim to recover the spectrum of the unknown circular convolution operator $A$. However, in the presence of outliers, even in only a few snapshots, existing approaches often struggle to recover the spectrum. We address this challenge by proposing a novel robust spectral recovery model in the presence of time-sparse corruptions. We propose a robust pipeline that lifts the problem to a sequence of robust low-rank Hankel recovery and completion tasks, followed by Prony-type spectral estimation. Numerical experiments confirm the accurate spectral recovery of the proposed approach and exhibit its superior robustness against state-of-the-art under various settings.

% %%%%%%%%%%%%%%%%%%%%%%%%%%%%%%%%%%%%%%%%%%%%%
% We study the robust spectral recovery problem for dynamical sampling on a finite cyclic grid. Given time snapshots of a fixed uniform spatial subsampling on $x_\ell=A^\ell f$, we aim to recover the spectra of the unknown circular convolution operator $A$. 
% In the presence of extreme outliers, even just in a few snapshots, existing approaches struggle with the recovery task. In this work, we lift the robust spectral recovery problem to a sequence of robust low-rank Hankel recovery tasks, followed by Prony-type spectral estimation. Experiments confirm that the proposed approach achieves accurate spectral recovery and shows superior robustness over the state-of-the-art.

%%%%%%%%%%%%%%%%%

% We allow an unknown, time-sparse subset of snapshots to be corrupted by arbitrarily large outliers,
% possibly along with additive Gaussian noise. Uniform subsampling yields an aliasing representation
% in which each frequency-channel time sequence is a mixture of exponentials, implying low-rank Hankel
% structure after Hankel lifting; time-sparse outliers become anti-diagonal sparse corruptions in the
% Hankel domain. Exploiting this structure, we propose a robust recovery pipeline that detects corrupted time indices via robust low-rank Hankel denoising and then performs Prony-type spectral estimation to
% recover the spectrum of $A$. Experiments  show accurate spectral recovery and clear improvements over Cadzow denoising.
\end{abstract}
 

\section{Introduction}
Spatiotemporal data acquisition problems often face a fundamental constraint: deploying dense
sensor arrays is expensive, while collecting repeated measurements in time is comparatively cheap.
\emph{Dynamical sampling} formalizes this time-spatial trade-off by studying recovery of    the underlying dynamics and/or an unknown initial state from coarse spatial samples collected across multiple time steps \cite{aldroubi2013dynamical,aldroubi2017dynamical,aldroubi2019frames,aceska2015multidimensional,cabrelli2020dynamical,aldroubi2020phaseless,beinert2023phase,huang2021robust,christensen2021survey,kummerle2022learning,gong2025randomized}.
The paradigm is motivated by applications in sensor networks, imaging and inverse problems \cite{hormati2009distributed,lu2009spatial,  reise2010reconstruction, ranieri2011sampling,  reise2012distributed}, where
the observed field evolves according to a known physical mechanism  but only a
subset of spatial locations can be measured repeatedly \cite{aldroubi2013dynamical}.
A prototypical example is the heat/diffusion equation, where the evolution acts as a strong
low-pass filter over time, enabling reconstruction from fewer spatial sensors when temporal
oversampling is available \cite{ranieri2011sampling}.

A central and practically relevant class of dynamics is spatially invariant evolution \cite{aldroubi2021sampling,huang2024convolutional},
where the evolution operator is a convolution. Convolution models arise naturally from discretizations of translation-invariant
PDEs (e.g., diffusion), from linear time-invariant systems, and from shift-invariant blurs in
signal and image processing. In this setting, the driving operator $A$ is completely characterized
by a short ``filter''  $a$, and the eigenvectors are the Fourier modes; hence $A$ is diagonalized by the discrete Fourier transform (DFT), with eigenvalues given by $\widehat a$.
This structure makes the model both expressive and computationally attractive: applying $A$ can be done rapidly via FFTs, and the unknown dynamics can be interpreted through
its spectrum $\widehat a$.

Within dynamical sampling, convolution operators have received special attention because they
allow \emph{system identification} from subsampled spatiotemporal measurements. In particular, \cite{aldroubi2016krylov,tang2017system} study identification of an unknown convolution filter $a$ and an unknown initial state $f$ from samples of the evolving states $\{f,Af,\dots,A^{L-1}f\}$ taken on a sublattice, and a Prony-type spectral method, with perturbation analysis, has been established. More broadly, the modern theory provides conditions under which the driving operator and/or the initial state can be recovered from time-spatial samples, clarifying when temporal oversampling compensates for spatial undersampling \cite{aldroubi2017dynamical}.

While existing dynamical sampling work has developed foundational identifiability conditions and
algorithms, robustness remains a key barrier in real deployments. Most analyses treat \emph{additive}
measurement noise (often stochastic) and study the stability of operator recovery under such perturbations.
For example, \cite{Aldroubi2018AdditiveNoise} analyzes the impact of additive random noise in the
finite-dimensional setting, including recovery of driving operators defined via real symmetric
convolution, and integrates denoising strategies to mitigate noise effects.
However, many sensing systems also experience \emph{gross, intermittent corruptions} that are better
modeled as \emph{sparse outliers in time}: a sensor may malfunction for a small number of time
instants, packets may be dropped or replaced by erroneous values, or a transient interference event
may corrupt a few snapshots. These sparse but potentially large-magnitude corruptions do not fit
well into the existing dynamical sampling models and can catastrophically bias spectral/Prony-type identification. 
To address the challenge posed by the outliers, this paper focuses on the following setting: Measurements are collected at fixed subsampled spatial locations across time, but a small subset of time snapshots are corrupted by outliers, no matter whether one or more sensors are impacted at the time. 

Later in \Cref{sec:method}, we show that the spectral recovery against time-sparse outliers can be lifted as a series of robust low-rank Hankel matrix recovery problems, with the ranks explicitly determined by spectral collisions induced by aliasing. Fortunately, robust low-rank Hankel recovery has been extensively exploited in various methodologies, such as nuclear norm minimization \cite{chen2014robust}, gradient descent \cite{cai2018spectral,cai2023structured}, Newton-like method \cite{cai2025hsnld}, Riemannian optimization \cite{cai2019fast}, and alternating projection \cite{zhang2019correction,cai2021asap}.

% We show that, for convolution-driven evolution under uniform spatial subsampling,
% Fourier-domain aliasing transforms the problem into estimating mixtures of exponentials in time.
% Consequently, the associated  Hankel matrices formed from the temporal sequences are
% low-rank, with ranks determined by spectral collisions induced by aliasing. This links dynamical
% sampling to the rich literature on Hankel-structured low-rank recovery and line spectral estimation.

% Low-rank Hankel structure has been extensively exploited in \emph{spectral compressed sensing}
% and super-resolution.... {\color{red}Hankel part}  \textcolor{blue}{A number of recent studies have focused on reconstructing low-rank Hankel matrices from a limited number of linear measurements, possibly corrupted by noise or outliers. These approaches include methods based on convex relaxations, such as nuclear norm minimization \cite{chen2014robust}, as well as a growing class of nonconvex methods that achieve significantly lower computational complexity while admitting rigorous theoretical guarantees, including gradient-based algorithms \cite{cai2018spectral,cai2023structured,cai2025hsnld}, Riemannian optimization inspired methods \cite{cai2019fast,cai2021asap}, and alternating projection schemes \cite{zhang2019correction}.}

\vspace{0.05in}
\noindent
\textbf{Contributions.} 
%\subsection{Contributions}
Our main contributions are threefold:
\begin{itemize}[leftmargin=*,itemsep=5pt, topsep=2pt, parsep=0pt, partopsep=0pt]
    \item We propose a novel robust spectral recovery model for dynamical sampling in the presence of time-sparse corruptions, which induce a sequence of robust low-rank Hankel recovery problems, with the low-rankness of the groundtruth guaranteed by \Cref{thm:lowrankproperty}. 
\item We propose a novel pipeline that leverages a sequence of robust Hankel recovery and completion algorithms to enable the stable Prony-style spectral recovery, with a provable upper bound on the fraction of tolerable outliers, grounded by \Cref{thm:Hankel_recovery}.
\item Numerical results under various settings show that the proposed approach tracks the true spectrum of the underlying convolution operator more accurately and robustly against the state-of-the-art, even in the presence of both outliers and additive noise.
\end{itemize}

\section{Notation and preliminaries}\label{sec:notation}
Let $d=mJ$ with integers $m\ge2$ and $J\ge1$. We work on cyclic groups
$\mathbb Z_d$ and $\mathbb Z_J$ (indices modulo $d$ and $J$). We use $\|\cdot\|_2$, $\|\cdot\|_F$,
and $\langle A,B\rangle=\mathrm{trace}(A^*B)$.

\paragraph{Circular convolution and Fourier transform.}
For $a,f\in\ell^2(\mathbb Z_d)$, define circular convolution
\[
(a*f)(k)=\sum_{t\in\mathbb Z_d}a(t)f(k-t).
\]
Let $\widehat{(\cdot)}$ denote the DFT on $\mathbb Z_d$. Then
$\widehat{a*f}(k)=\widehat a(k)\widehat f(k)$, so the convolution operator $Af:=a*f$ is
diagonalized by the DFT with spectrum $\{\widehat a(k)\}_{k\in\mathbb Z_d}$
\cite{aldroubi2013dynamical,tang2017system}.

\paragraph{Uniform subsampling.}
The uniform subsampling operator $S_m:\ell^2(\mathbb Z_d)\to\ell^2(\mathbb Z_J)$ is
\[
(S_m z)(j)=z(mj),\qquad j\in\mathbb Z_J,
\]
and satisfies the  identity
\begin{equation}\label{eq:aliasing}
\widehat{(S_m z)}(j)=\frac{1}{m}\sum_{n=0}^{m-1}\widehat z(j+nJ),\qquad j\in\mathbb Z_J,
\end{equation}
where $\widehat z$ on the right is the DFT on $\mathbb Z_d$
\cite{aldroubi2013dynamical,tang2017system}.

\paragraph{Hankel lifting}
Given a scalar sequence $\{s_\ell\}_{\ell=0}^{L-1}$ and $K$ with $2K\le L$, define the $K\times K$
Hankel matrix
\begin{equation}\label{eq:Hankel}
  H_K(s):=\bigl[s_{p+q}\bigr]_{p,q=0}^{K-1}.
\end{equation}
We write $ \mathbb{H}_K$ for the space of all $K\times K$ Hankel matrices.
\section{Problem Setup}\label{sec:setup}

\subsection{Dynamical sampling with time-sparse corruptions}
Let $A$ be the unknown convolution operator $Af=a*f$ on $\ell^2(\mathbb Z_d)$. Given an unknown initial state $f\in\ell^2(\mathbb Z_d)$, the orbit is
\[
x_\ell = A^\ell f,\quad \ell\in\{0,\cdots,L-1\} \text{~with~} [L]:=\{1,\cdots,L\}.
\]
We observe uniform subsamples
\begin{equation}\label{eq:meas}
y_\ell = S_m x_\ell \in \ell^2(\mathbb Z_J),\qquad \ell\in \{0,\cdots,L-1\},
\end{equation}
where $S_m$ is defined in \Cref{sec:notation}. We focus on \emph{time-sparse outliers}:
there exists an unknown set $\Omega_{\mathrm{out}}\subset\{0,\cdots,L-1\}$ with
$|\Omega_{\mathrm{out}}|\ll L$ such that
\begin{equation}\label{eq:time_outlier_model}
\widetilde y_\ell = y_\ell + e_\ell,\quad
e_\ell=0,\quad \text{for }\ell\notin\Omega_{\mathrm{out}},
\end{equation}
and the nonzero $e_\ell$ can have arbitrary magnitudes (gross snapshot corruptions).

\subsection{Low-rank Hankel structure in Fourier-domain}
Let's define
\[
s_\ell(j):=\widehat y_\ell(j),\quad \widetilde s_\ell(j):=\widehat{\widetilde y_\ell}(j),
\quad j\in\mathbb Z_J.
\]
By \eqref{eq:aliasing} and $\widehat{A^\ell f}(k)=\widehat a(k)^\ell\widehat f(k)$, each
$\{s_\ell(j)\}_{\ell\ge0}$ is a finite mixture of exponentials in $\ell$, hence the Hankel matrix
$H_K(j):=\mathcal H_K(\{s_\ell(j)\})$ is low rank (Theorem~\ref{thm:lowrankproperty}).
Moreover, \eqref{eq:time_outlier_model} implies
\[
\widetilde H_K(j):=\mathcal H_K(\{\widetilde s_\ell(j)\}) = H_K(j) + O_K(j),
\]
where $O_K(j)\in\mathbb{H}_K$ has nonzero anti-diagonals only at indices in $\Omega_{\mathrm{out}}$.

\subsection{Goals}
Given the corrupted dynamical samples $\{\widetilde y_\ell\}_{\ell=0}^{L-1}$, our goals are:
\begin{itemize}[leftmargin=*, itemsep=5pt, topsep=2pt, parsep=0pt, partopsep=0pt]
\item   \textbf{Robust Hankel recovery:} recover the low-rank Hankel matrices $\{H_K(j)\}_{j\in\mathbb{Z}_J}$
and identify the outlier set $\Omega_{\mathrm{out}}$ from the structured observations
$\{\widetilde H_K(j)\}$;

\item  \textbf{Robust spectral recovery:} estimate the aliased spectra
$\{\widehat a(j+nJ)\}_{n=0}^{m-1}$ for each $j$, and thereby recover the spectrum (and, when
identifiable, the filter $a$) of the driving operator $A$.
\end{itemize}

\section{Methodology}\label{sec:method}

This section addresses the two goals in \Cref{sec:setup}: (i) robust recovery of the clean
Hankel matrices and the outlier time indices, and (ii) recovery of the spectrum of the driving
convolution operator $A$.

%------------------------------------------------
\subsection{Low-rank Hankel structure}\label{subsec:lowrank}
We first recall the key structural fact: after uniform subsampling, each Fourier channel
produces a mixture of exponentials in time, which implies low-rank Hankel structure.

\begin{theorem}[Low-rank dynamical Hankel matrices]\label{thm:lowrankproperty}
Let $d=mJ$ with $d$ odd, and let $a,f\in \ell^2(\mathbb Z_d)$. Fix $j\in\mathbb Z_J$ and define
\[
y_\ell=S_m(a^\ell*f),\quad s_\ell(j):=\widehat y_\ell(j),\quad  \ell\in\{0,\cdots,L-1\}.
\]
For an integer $K\ge1$ with $2K\le L$, define the Hankel matrix $$H_K(j):=\bigl[s_{p+q}(j)\bigr]_{p,q=0}^{K-1}.$$
Let $r_j$ be the number of distinct
values among $\{\widehat a(j+nJ)\}_{n=0}^{m-1}$. Then for any $K\ge r_j$,
\[
\rank(H_K(j))\le r_j,
\]
and for generic $f$ equality holds. In particular, if $\widehat a$ is real symmetric and strictly
decreasing on $\bigl[0,\frac{d-1}{2}\bigr]$, then
\[
r_j=
\begin{cases}
\frac{m+1}{2}, & j=0;\\
m, & j> 0.\\
\end{cases}
\]
\end{theorem}

% \noindent\emph{Proof sketch.}
% By the aliasing identity \eqref{eq:aliasing} and $\widehat{a^\ell*f}(k)=\widehat a(k)^\ell\widehat f(k)$,
% \[
% s_\ell(j)=\frac{1}{m}\sum_{n=0}^{m-1}\widehat a(j+nJ)^\ell\,\widehat f(j+nJ),
% \]
% which is a linear combination of at most $r_j$ exponentials in $\ell$.
% Thus $H_K(j)$ admits a Vandermonde factorization $H_K(j)=V\,\text{diag}(\alpha)\,V^\top$ and is rank $\le r_j$.
% \hfill $\square$

\medskip
Theorem~\ref{thm:lowrankproperty} yields a \emph{rank separation} between $j=0$ and $j>0$ under
mild spectral monotonicity, which we exploit for robust outlier localization.

%------------------------------------------------
\subsection{Robust Hankel recovery with time-sparse outliers}\label{subsec:robust_hankel}
Recall the time-sparse corruption model \eqref{eq:time_outlier_model}.
After DFT on $\mathbb Z_J$ and Hankel lifting, for each $j\in\mathbb Z_J$,
\[
\widetilde H_K(j)=H_K(j)+O_K(j),
\]
where $O_K(j)\in\mathbb H_K$ is supported only on anti-diagonals indexed by the outlier set
$\Omega_{\mathrm{out}}$ (equivalently, few corrupted time indices).

Let $\mathcal I:=\{0\}\cup[2K-2]$ be the anti-diagonal index set.
We first detect corrupted anti-diagonals using the  channel $j=0$, and then complete the remaining channels using the detected clean indices.

\vspace{0.05in}
\noindent
\textbf{Stage I: Reference Hankel matrix at $j=0$.} Under the assumptions of \Cref{thm:lowrankproperty}, the Hankel matrix $\bm{H}_K(0)$ has the smallest rank $\frac{m+1}{2}$, making it well suited for outlier detection, since lower rank Hankel matrices can tolerate a larger fraction of sparse outliers \cite{chen2014robust}. Accordingly, the outlier detection task can be formulated as the following optimization problem:
\begin{equation} \label{task:outlier}
    \begin{split}
        \min_{H,O\in\mathbb {H}_K}~&\frac{1}{2}\|\widetilde{H}_K(0)-H-O\|_F^2\cr
\text{s.t.}
\quad &\mathrm{rank}(H)=\tfrac{m+1}{2},
\;O\ \text{is sparse},
    \end{split}
\end{equation}
To solve \eqref{task:outlier}, we employ the \emph{Accelerated Structured Alternating Projections (ASAP)} algorithm proposed in \cite{cai2021asap}, which is specifically designed for outlier detection of low-rank Hankel matrices. The recovered sparse component $\widehat{O}$ identifies the corrupted anti-diagonals of the Hankel matrix and thereby provides an estimate $\widehat\Omega_{\mathrm{out}}$ of the outlier support. 

\vspace{0.05in}
\noindent
\textbf{Stage II: Hankel matrix completion for $j> 0$.} Once the outlier support has been estimated from the reference frequency $j=0$,
we turn to the recovery of the remaining Hankel matrices.
For $j>0$, we remove the detected outlier indices and solve
a low rank Hankel completion problem:
\begin{equation} \label{task:completion}
    \begin{split}
        \min_{H\in\mathbb H_K}~
&\frac{1}{p}
\langle \mathcal P_{\Omega^c}
(H-\widetilde{H}_K(j)),H-\widetilde{H}_K(j)
\rangle \cr
\text{s.t.}\quad &\mathrm{rank}(H)=m,
    \end{split}
\end{equation}
where $\Omega$ corresponds to anti-diagonals indexed by $\widehat\Omega_{\mathrm{out}}$, and $p:=\frac{\Omega^c}{L}$ is the rate of uncorrupted observation. The remaining task is therefore to solve \eqref{task:completion}. To this end, we employ the \emph{Hankel Structured Newton-Like Descent (HSNLD)} algorithm proposed in \cite{cai2025hsnld}, which is designed for robust low-rank Hankel matrix completion.

%------------------------------------------------
\subsection{Spectrum recovery from cleaned  Hankel matrices}
\label{sec:spectrum_recovery}
We now describe how to recover the spectrum of the driving operator $A$ once the time-domain outliers have been removed via robust Hankel recovery.
Throughout this subsection we focus on the convolution model
$A x = a * x$  with real-symmetric filter $a$, so that $A$ is
diagonalized by the DFT and
$\sigma(A)=\{\widehat a(k)\}_{k\in\mathbb Z_d}$. According to previous analysis,
each fixed frequency $j\in\mathbb Z_J$ yields a scalar time sequence
$\{s_\ell(j)\}_{\ell\ge 0}$ that is a linear combination of exponentials in $\ell$, i.e., $s_\ell(j)=\frac{1}{m}\sum_{n=0}^{m-1}\widehat a(j+nJ)^\ell\,\widehat f(j+nJ)$. 
Consequently, $\{s_\ell(j)\}$ satisfies a short linear recurrence whose
characteristic polynomial has roots equal to the \emph{distinct} values among
$\{\widehat a(j+nJ)\}_{n=0}^{m-1}$ (cf.~\cite{Aldroubi2018AdditiveNoise,aldroubi2016krylov}).

\vspace{0.08in}
\noindent
\textbf{Step 0: Input from Hankel recovery.} 
From Section~\ref{subsec:robust_hankel}, we obtain denoised sequences
$\{\widehat s_\ell(j)\}_{\ell=0}^{L-1}$ for each $j\in\mathbb Z_J$
(or equivalently denoised Hankel matrices $\widehat H_K(j)$).
In what follows we write $s_\ell(j)$ for the cleaned/denoised sequence.

\vspace{0.08in}
\noindent
\textbf{Step 1: Estimate the model order $r_j$.}
Let $r_j:=\rank(H_K(j))$ (in the noiseless case, $r_j$ equals the number of
distinct values among $\{\widehat a(j+nJ)\}_{n=0}^{m-1}$; see
Theorem~\ref{thm:lowrankproperty}).
In practice, we estimate $r_j$ by a numerical rank test on $\widehat H_K(j)$.

\vspace{0.08in}
\noindent
\textbf{Step 2: Compute the annihilating polynomial.}
For each $j$, define the coefficient vector
$c(j)=(c_0(j),\dots,c_{r_j-1}(j))^\top$ by fitting the recurrence
\begin{equation*}
s_{k+r_j}(j)+\sum_{\ell=0}^{r_j-1} c_\ell(j)\,s_{k+\ell}(j)=0,
  k\in\{0\}\cup[K-r_j-1],
\end{equation*}
which is the finite-dimensional analogue of the minimal-polynomial relation
used in \cite{Aldroubi2018AdditiveNoise}.
This can be written as a Hankel linear system
\begin{equation}\label{eq:linpred_system}
 [H_K(j)]_{:,1:r_j}c(j)=[H_K(j)]_{:,r_j+1}.
\end{equation}
% \begin{equation}\label{eq:linpred_system}
% \underbrace{\begin{bmatrix}
% s_0(j) & s_1(j) & \cdots & s_{r_j-1}(j)\\
% s_1(j) & s_2(j) & \cdots & s_{r_j}(j)\\
% \vdots & \vdots & \ddots & \vdots\\
% s_{K-r_j-1}(j) & s_{K-r_j}(j) & \cdots & s_{K-2}(j)
% \end{bmatrix}}_{=:~T_j\in\mathbb C^{(K-r_j)\times r_j}}
% c(j)
% =
% -\underbrace{\begin{bmatrix}
% s_{r_j}(j)\\
% s_{r_j+1}(j)\\
% \vdots\\
% s_{K-1}(j)
% \end{bmatrix}}_{=:~b_j\in\mathbb C^{K-r_j}}.
% \end{equation}
If $K>2r_j$, we solve \eqref{eq:linpred_system} in the least-squares sense for
robustness to residual denoising error.

We then form the monic polynomial
\begin{equation*}\label{eq:pj_def}
p_j(\lambda)=\lambda^{r_j}+\sum_{\ell=0}^{r_j-1} c_\ell(j)\,\lambda^\ell.
\end{equation*}
 
\vspace{0.08in}
\noindent
\textbf{Step 3: Recover the  spectral values.}
Let $R(j)$ be the set of roots of $p_j$. In the noiseless case, 
\begin{equation}\label{eq:root_set}
R(j)= \{\lambda_{j,t}\}_{t=1}^{r_j}=\cup_{n=0}^{m-1}\{\widehat a(j+nJ)\}.
\end{equation}
After robust Hankel recovery, $R(j)$ provides a stable estimate of these values.

% \paragraph{Step 4 (Assemble the spectrum and (optionally) the filter).}
% Finally, we estimate the spectrum of $A$ by
% \begin{equation}\label{eq:spectrum_union}
% \widehat{\sigma(A)}=\bigcup_{j=0}^{J-1} R(j).
% \end{equation}
% If one additionally wants to recover the full DFT vector $\widehat a(k)$ (hence $a$),
% one needs to \emph{label} which root in $R(j)$ corresponds to each $k=j+nJ$.
% This labeling is identifiable under mild structural assumptions on $\widehat a$
% (e.g., real-symmetry and monotonicity on $[0,(d-1)/2]$), after which $a$ is obtained
% by an inverse DFT.

%\section{Main  Results}
\subsection{Recovery guarantees of low rank Hankel matrix} 

To ensure a low-rank matrix is recoverable from partial observations, we first recall the incoherence condition, which is widely used in matrix completion \cite{candes2012exact}. For Hankel matrices, this condition can be defined in terms of Vandermonde decomposition. Note 
the Hankel entries satisfy $ H_K(j)_{p,q}=s_{p+q}(j)$ with
\begin{align*}
   s_{p+q}(j) 
   =\frac{1}{m}\sum_{n=0}^{m-1}\widehat a(j+nJ)^{p+q}\,\widehat f(j+nJ)
=\sum_{t=1}^{r_j}\alpha_{j,t}\,\lambda_{j,t}^{\,p+q}.
\end{align*}
Therefore,  $$H_K(j)=V_K(j)D(j)V_K(j)^{\top}$$
with $V_K(j)=[\lambda_{j,t}^{p}]_{p=0,t=1}^{K-1,r_j}$ and $D(j)=\mathrm{diag}(\alpha_{j,1},\dots,\alpha_{j,r_j})$. Then, the incoherence condition for the underlying Hankel matrix $H_K(j)$ is defined as follows. 
\begin{definition}[Incoherence \cite{chen2014robust,cai2018spectral}] \label{def:incoherence}
	The Hankel matrix $H_K(j)$ with the Vandermonde decomposition $H_K(j)=V_K(j) D(j) V_K(j)^{\top}$ is said to be $\mu$-incoherent if the smallest singular values
$\sigma_{\text{min}} (V_K^{\top} V_K ) \ge \frac{K}{\mu}$.
\end{definition}
Let $\kappa_j$ be the condition number of $H_K(j)$, $j\in \mathbb{Z}_J$, denote $\kappa :=\max_{j}\{\kappa_j\}$. Then, we have the following recovery guarantee.

\begin{theorem}[Recovery guarantee for Hankel matrices]\label{thm:Hankel_recovery} 
Suppose $\{H_K(j)\}_{j\in\mathbb{Z}_J}$ are $\mu$-incoherent, and it holds  
\begin{align*}
|\Omega_{\mathrm{out}}| 
\le&\min\left\{L-\mathcal{O}(\mu^2 m^2\kappa^4\log L\log\frac{1}{\varepsilon}),\right. \cr
&\qquad\left.\mathcal{O}(\frac{4L}{\mu^2 (m+1)^2\kappa^2}) \right\}.%\cr |\Omega_{\mathrm{out}}| &\lesssim \mathcal{O}(\frac{4T}{\mu^2 (m+1)^2\kappa^2}). 
\end{align*}
with $L\gg\log(1/\varepsilon)$ . Then, with probability at least $1-\frac{J}{L}$, the solutions to \eqref{task:outlier} and \eqref{task:completion} obtained via ASAP and HSNLD yield $\varepsilon$-accurate estimates $\{\hat{H}_K(j)\}_{j\in\mathbb Z_J}$ of $\{H_K(j)\}_{j\in\mathbb Z_J}$, that is,
$$\|\hat{H}_K(j)-H_K(j)\|_{F}\le \varepsilon,\quad \forall j\in \mathbb Z_J.$$
\end{theorem}

\section{Numerical Evaluation} \label{sec:numerical}
We evaluate the proposed robust Hankel-Prony pipeline for spectral recovery in convolutional dynamical sampling under \emph{time-sparse outliers} and \emph{additive Gaussian noise}.
Throughout, the convolution operator is diagonalized by the DFT, and its spectrum is denoted by
$\widehat a\in\mathbb{R}^{n}$ (real-symmetric in our construction).
We report results in terms of (i) spectrum plots and (ii) spectral error/SNR metrics.

\subsection{Data generation and corruption model}
We work on a cyclic grid $\mathbb{Z}_d$ with $d$ odd and generate a real-symmetric spectrum
$\widehat a$ by drawing $\widehat a(0)=1$, sampling i.i.d.\ entries on $\{1,\dots,(d-1)/2\}$,
and mirroring them to enforce $\widehat a(k)=\widehat a(d-k)$.
We simulate convolutional dynamics as following:
\[
x_{\ell+1}=\mathcal{F}^{-1}\bigl(\widehat a\odot \mathcal{F}x_{\ell}\bigr),\quad \ell=0,1,\dots,L-1,
\]
with $L=300$ starting from a random $x_0$. Uniform spatial subsampling uses $J=d/m$ sensors at locations $\{0,m,2m,\dots,(J-1)m\}$, producing
$y_\ell=S_m x_\ell\in\mathbb{R}^{J}$.
We then inject:
\begin{itemize}[leftmargin=*, itemsep=0pt, topsep=2pt, parsep=0pt, partopsep=0pt]
\item  \textbf{Time-sparse outliers:} Choose an outlier index set $\Omega_{\mathrm{out}}\subset\{0,\dots,L-1\}$
uniformly at random with $|\Omega_{\mathrm{out}}|=\lfloor \alpha L\rfloor$.
For each $\ell\in\Omega_{\mathrm{out}}$, we corrupt the snapshot by adding a perturbation
$e_\ell\in\mathbb{R}^J$ whose entries are drawn i.i.d. from a uniform distribution with scale
proportional to the snapshot magnitude:
$(e_\ell)_j \sim \mathrm{Unif}\!\Bigl(\,-c\,\mu_\ell,\; c\,\mu_\ell\,\Bigr)$, $j\in[J]$,
where $\mu_\ell := \frac{1}{J}\|y_{\ell}\|_1$. 
For $\ell\notin\Omega_{\mathrm{out}}$, we set $e_\ell=0$.

\item \textbf{Additive Gaussian noise:} Add i.i.d.~Gaussian noise $g_\ell$ with standard deviation
$\sigma\in\{10^{-3},10^{-5},10^{-7},10^{-9}\}$.
\end{itemize}
The observed samples are $\widetilde y_\ell = y_\ell + e_\ell + g_\ell$.

\subsection{Methods compared}
\begin{enumerate}[leftmargin=*, itemsep=5pt, topsep=2pt, parsep=0pt, partopsep=0pt]
\item \textbf{Proposed (robust Hankel + completion + Prony).}
We apply robust Hankel denoising on the reference frequency channel (the lowest-rank Hankel case),
estimate the corrupted time set, complete the remaining channels using Hankel completion, and finally
recover the spectral values using the annihilating-filter/Prony step.
\item \textbf{Cadzow baseline (Cadzow + Prony).}
We use  the state-of-the-art \emph{Cadzow denoising} method  (alternating projections onto the low-rank and Hankel constraints) as our baseline, which is applied independently to each channel and followed by the same Prony step.
This route is a standard and widely used stabilization heuristic for Prony/Hankel-based spectral estimation and has also
appeared in the dynamical sampling literature as a denoising pre-step (often referred to as \emph{Cadzow-like denoising});
see \cite{Aldroubi2018AdditiveNoise,tang2017system,cadzow2003high,gillard2010cadzow}.
\end{enumerate}

%\vspace{-0.08in}
\subsection{Evaluation metrics}
%\vspace{-0.05in}

Let $s=\{\hat{a}(k)\}_{k=0}^{d-1}$ and $\widehat s$ be the corresponding estimate. We will report:
\begin{itemize}[leftmargin=*,itemsep=5pt, topsep=2pt, parsep=0pt, partopsep=0pt]
\item \textbf{Relative spectral error:}\quad
$\mathrm{RE} := \|\widehat s-s\|_2/\|s\|_2$.
\item \textbf{Spectral SNR (dB):}
$\mathrm{SNR}_{\mathrm{spec}} := -20\log_{10}\mathrm{RE}$.
\end{itemize}

To quantify measurement corruption strength in the observation domain,
let $Y=[y_0,\dots,y_{L-1}]$, $E=[e_0,\dots,e_{L-1}]$, $G=[g_0,\dots,g_{L-1}]$, and
$\widetilde Y = Y+E+G$. We will report  
% $
% \mathrm{SNR}_{\mathrm{Outlier}} := 20\log_{10}(\frac{\|Y\|_F} {\|E\|_F})$ and $
% \mathrm{SNR}_{\mathrm{Gauss}} := 20\log_{10}\frac{\|Y\|_F}{\|G\|_F})$. 
%$\mathrm{SNR}_{\mathrm{Outlier}}$ and $\mathrm{SNR}_{\mathrm{Gauss}}$. 
$
\mathrm{SNR}_{\mathrm{Outlier}} := 20\log_{10}(\|Y\|_F /\|E\|_F)$ and $
\mathrm{SNR}_{\mathrm{Gauss}} := 20\log_{10}(\|Y\|_F/\|G\|_F)$. %\mathrm{SNR}_{\mathrm{tot}} := 20\log_{10}\frac{\|Y\|_F}{\|E+G\|_F}.

\subsection{Spectrum recovery under mixed corruptions}

\begin{figure}[t]
  \centering
%   \subfloat[$\sigma=10^{-4}$]{\includegraphics[width=.33\linewidth]{Fig/fig_spectrum_sigma_1e-04.png}} \hfill
% \subfloat[$\sigma=10^{-6}$]{\includegraphics[width=.33\linewidth]{Fig/fig_spectrum_sigma_1e-06.png}}\hfill
% \subfloat[$\sigma=10^{-8}$]{\includegraphics[width=.33\linewidth]{Fig/fig_spectrum_sigma_1e-08.png}}
\subfloat[$\sigma=10^{-3}$]{\includegraphics[width=.49\linewidth]{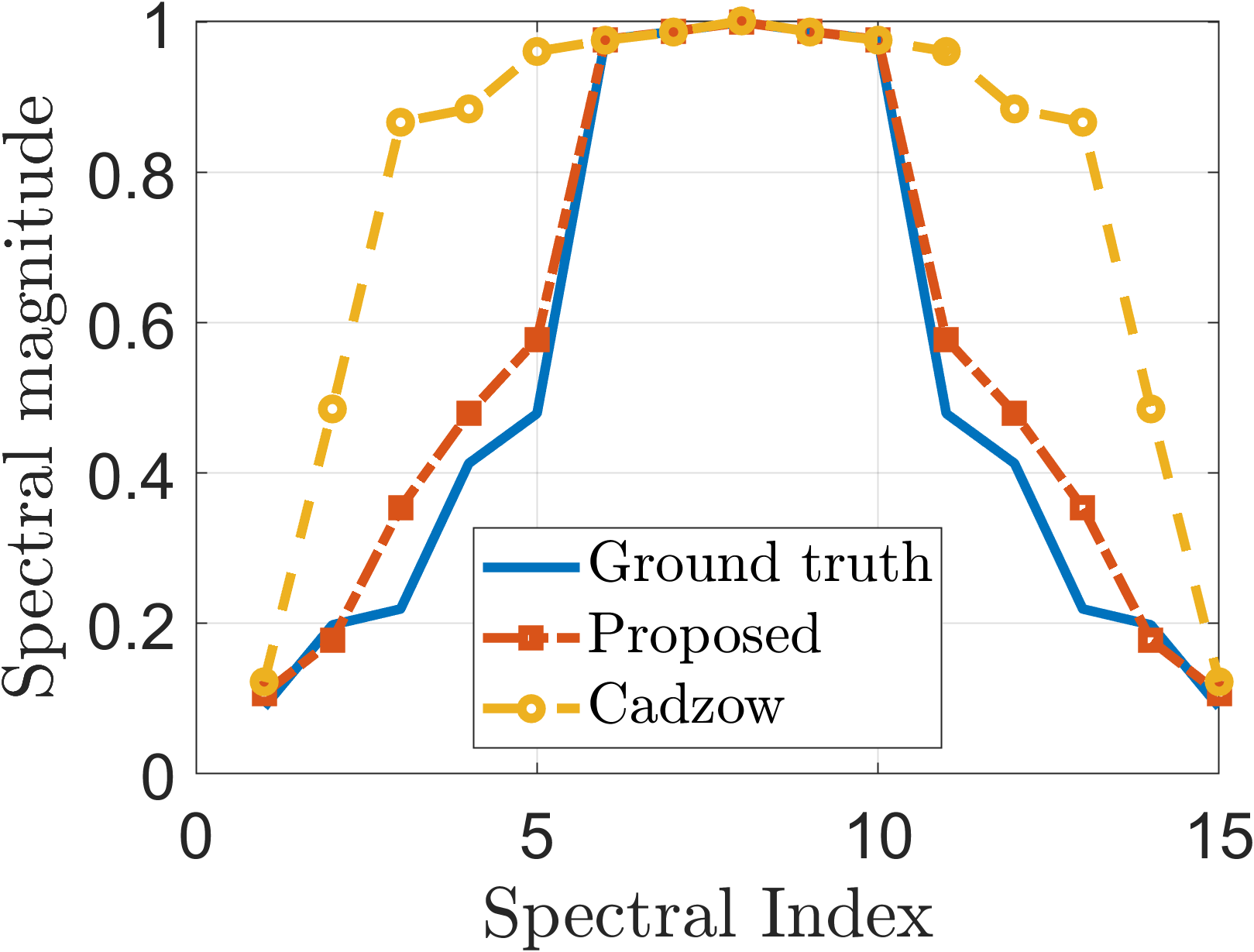}} \hfill
\subfloat[$\sigma=10^{-5}$]{\includegraphics[width=.49\linewidth]{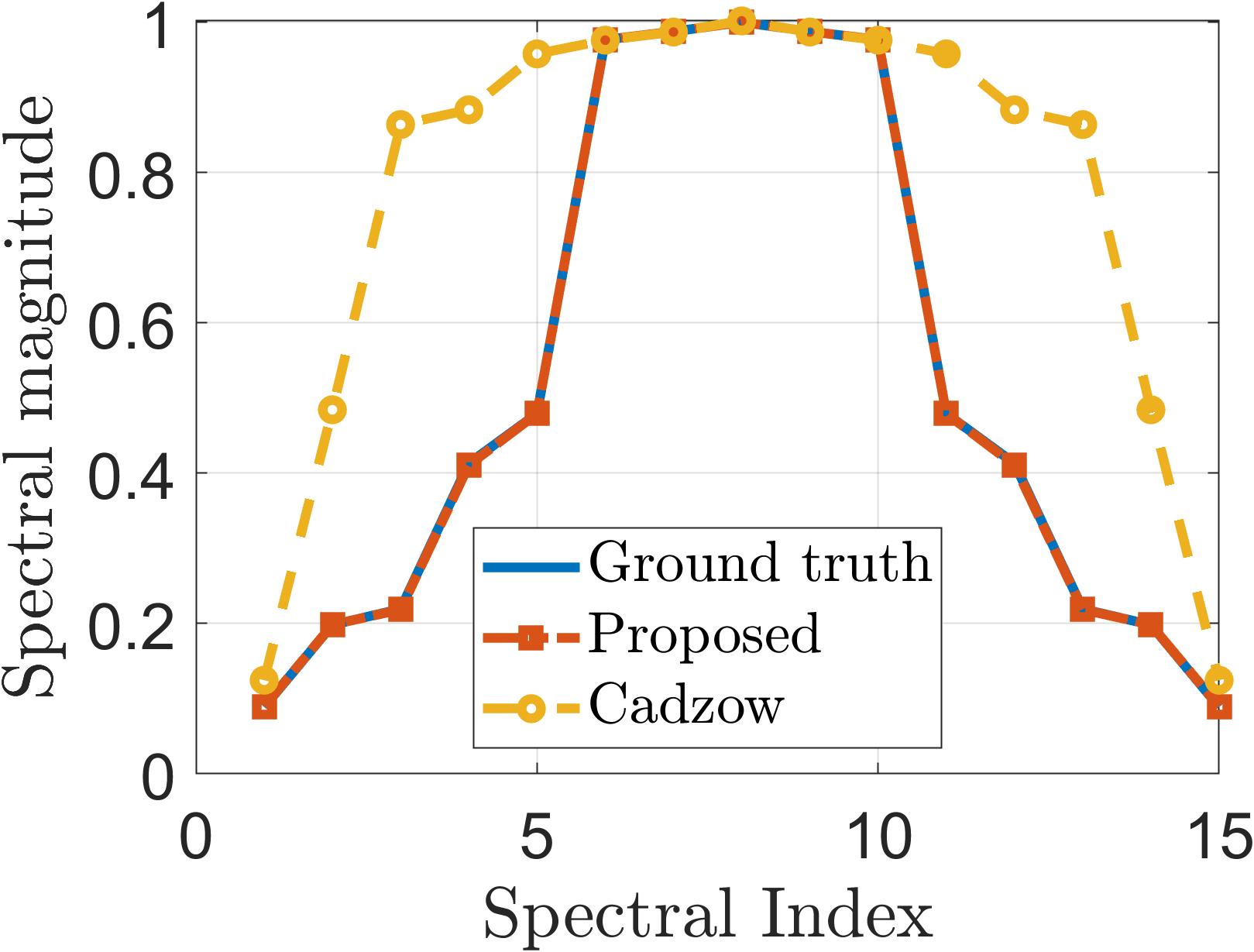}}\\
\subfloat[$\sigma=10^{-7}$]{\includegraphics[width=.49\linewidth]{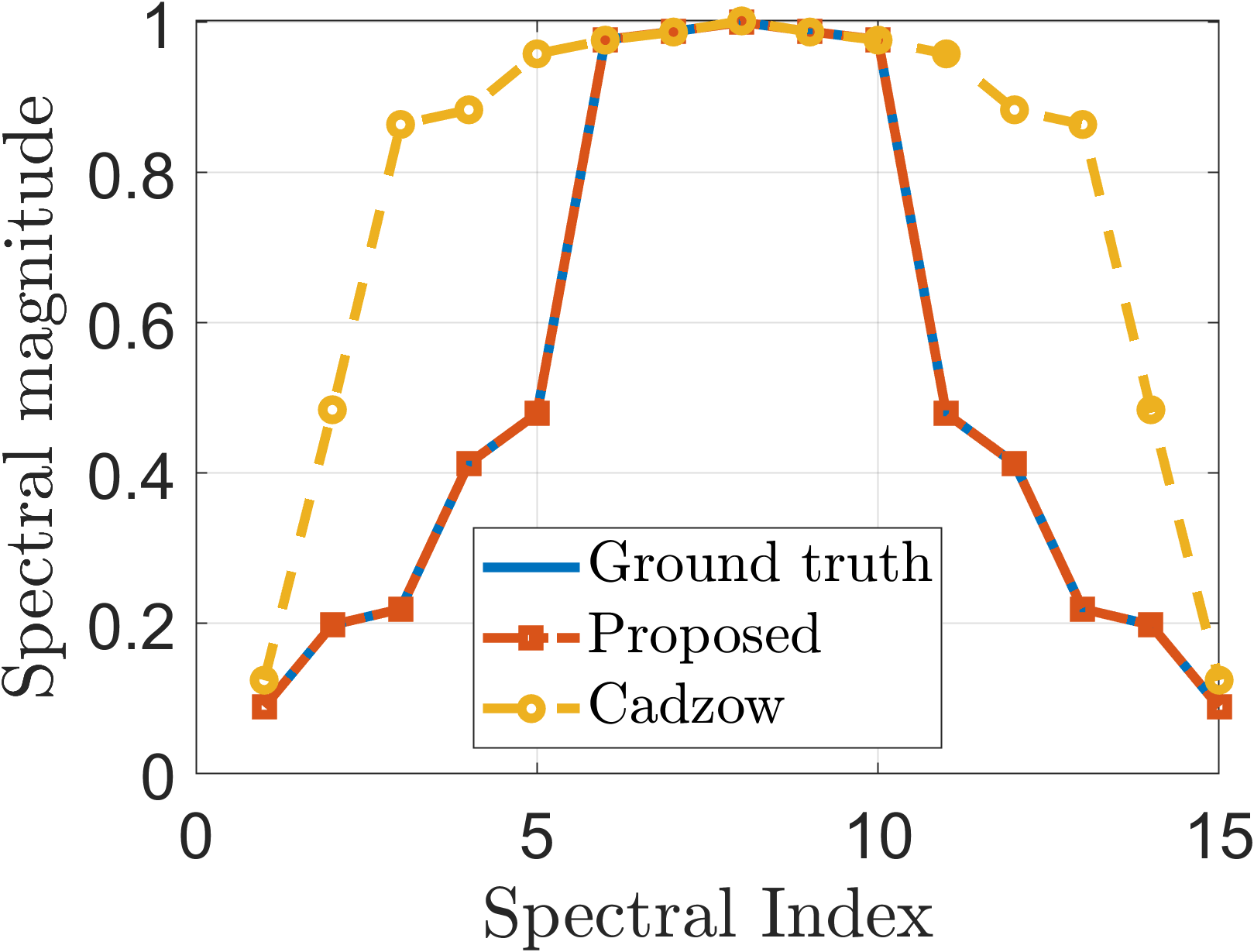}}\hfill
\subfloat[$\sigma=10^{-9}$]{\includegraphics[width=.49\linewidth]{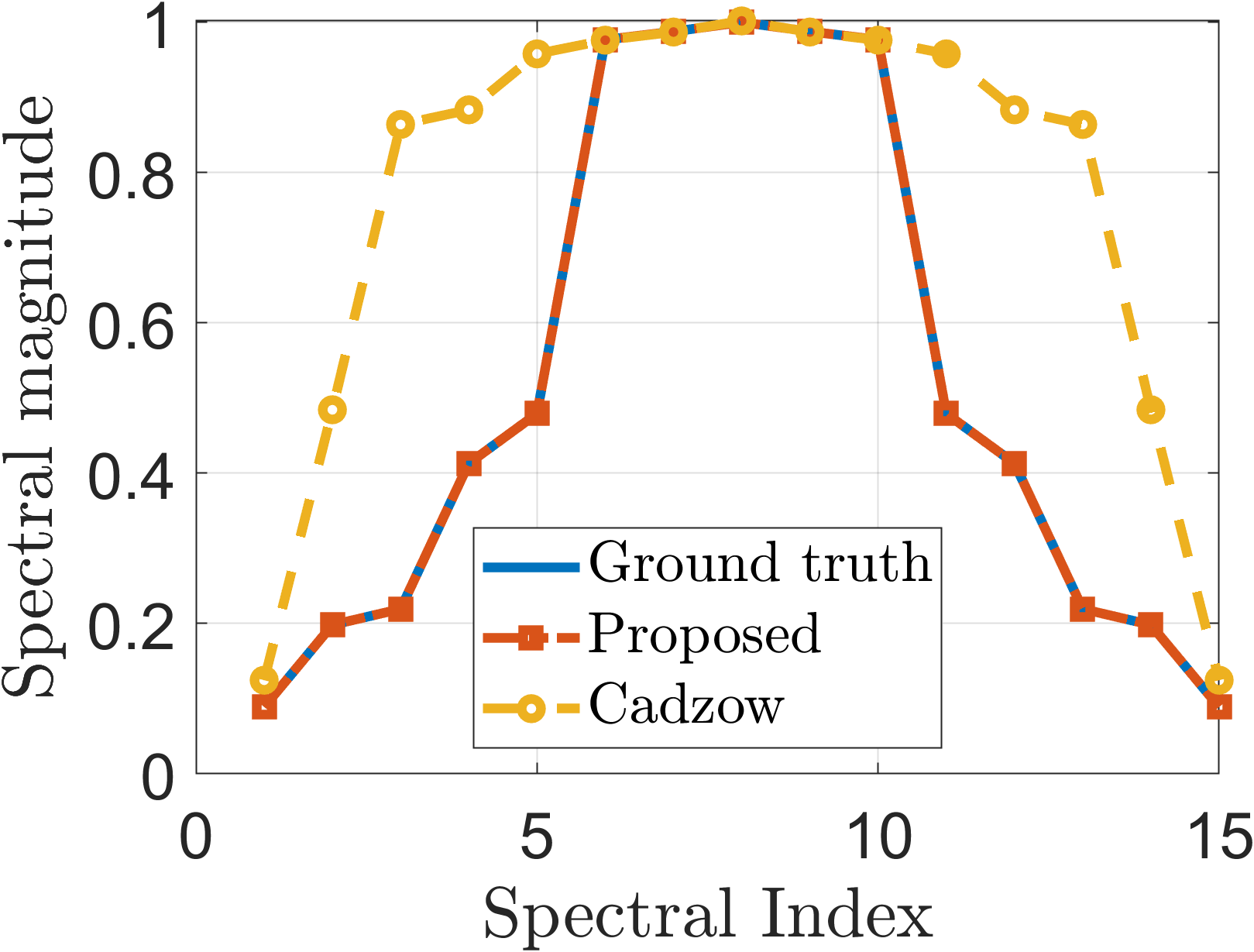}}
  \caption{Spectral recovery under   time-sparse outliers (rate $\alpha=5\%$) and additive Gaussian noise at three noise levels.
  Each panel plots the spectral magnitude $\widehat a(k)$ (ground truth) and the recovered spectra from
  \textbf{Proposed} (robust Hankel denoising + Prony) and \textbf{Cadzow} (Cadzow denoising + Prony). As the Gaussian noise level decreases, the proposed method increasingly matches the true spectrum and resolves the correct spectral peaks,
whereas the Cadzow baseline remains biased due to its sensitivity to sparse, structured corruptions.}
  \label{fig:spectrum_three_sigmas_row}
\end{figure}
 \begin{table}[ht]
\centering
%\footnotesize
\caption{Signal-to-noise ratios (dB) for the measurements and the recovered spectra under mixed sparse outliers and   Gaussian noise.   As $\sigma$ decreases, the proposed method continues to improve and tracks the true spectrum increasingly accurately, whereas the Cadzow baseline exhibits limited gains due to its sensitivity to sparse, structured outliers.}
\label{tab:snr_results}
\setlength{\tabcolsep}{9pt}
\renewcommand{\arraystretch}{1.15}
\resizebox{\linewidth}{!}{
\begin{tabular}{ccc|cc}
\toprule
$\sigma$&$\mathrm{SNR}_{\text{Gauss}}$ & $\mathrm{SNR}_{\text{Outlier}}$ & $\mathrm{SNR}_{\text{Proposed}}$ & $\mathrm{SNR}_{\text{Cadzow}}$ \\
\midrule
$10^{-3}$ & ~42.8783 &7.7888 &     ~19.4622  & 4.8494
\\      
$10^{-5}$& ~82.7459  &7.7888   &   ~56.6587  &   4.8893\\
  $10^{-7}$& 123.2371  &   7.7888    & ~90.4528 &    4.8886\\
  $10^{-9}$& 162.6056&  7.7888       &128.4645   &  4.8886\\
\bottomrule
\end{tabular}
}
\end{table}
We fix $(d,m,J,L)=(15,3,5,300)$ and a time-outlier rate $\alpha=5\%$, and test several Gaussian noise levels.
For each $\sigma$, we plot the recovered spectrum against the ground truth and record the corresponding SNRs.

\Cref{fig:spectrum_three_sigmas_row} shows representative recovered spectra at three noise levels
($\sigma=10^{-3},10^{-5},10^{-7},10^{-9}$).
Across all cases, our method closely tracks the ground truth despite time-sparse outliers,
whereas the Cadzow baseline is noticeably biased under sparse, potentially large corruptions.
The SNR summaries in \Cref{tab:snr_results} corroborate this trend:
as $\sigma$ decreases (i.e., the Gaussian noise weakens and the measurement SNR increases),
the recovered-spectrum SNR of our method improves substantially, while the Cadzow-based method
remains much lower.

\subsection{Robustness versus corruption rate}
\begin{figure}[th]
  \centering
   %\subfloat[$c=1$]
{\includegraphics[width=.49\linewidth]{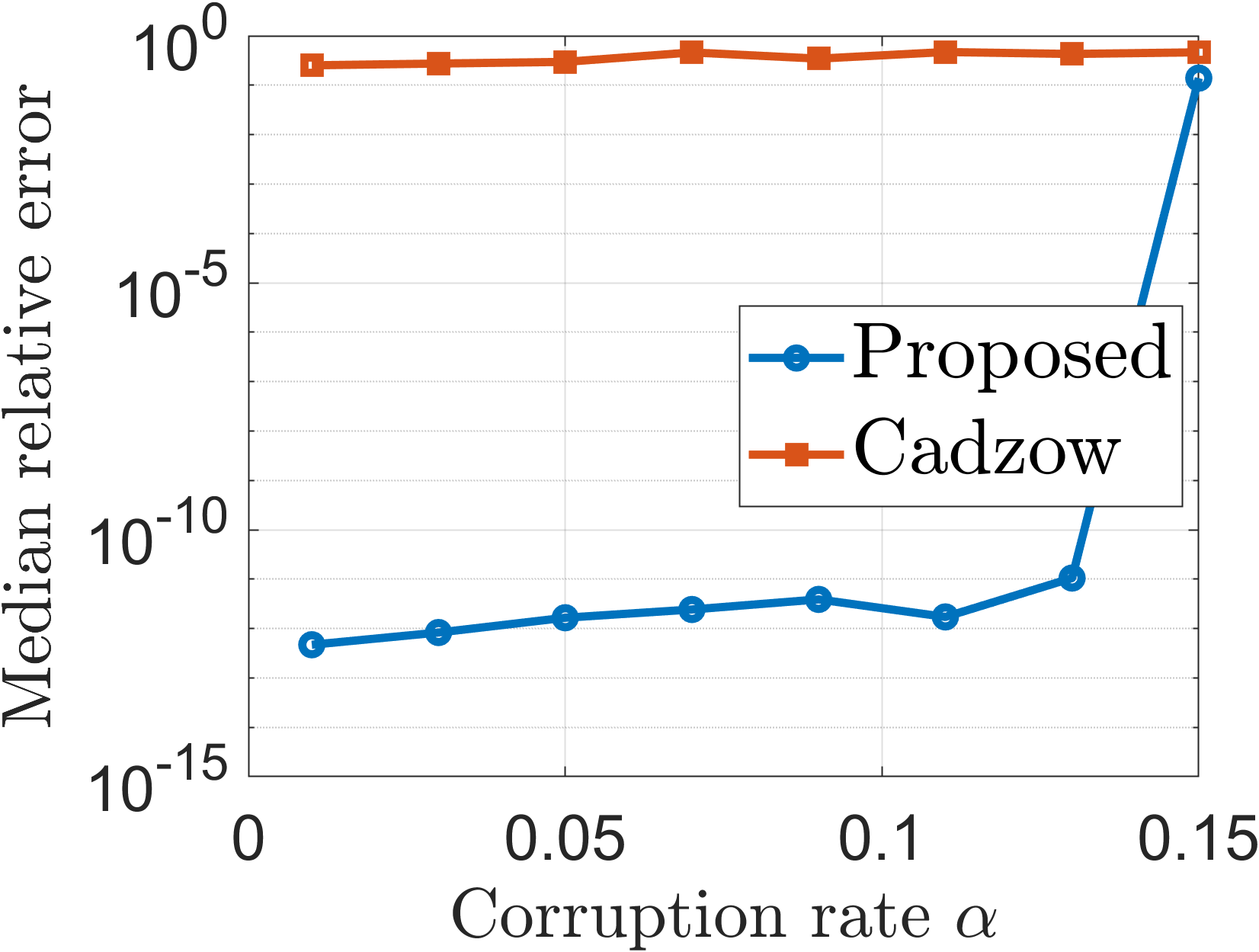}} 
%\subfloat[$c=5$]
{\includegraphics[width=.49\linewidth]{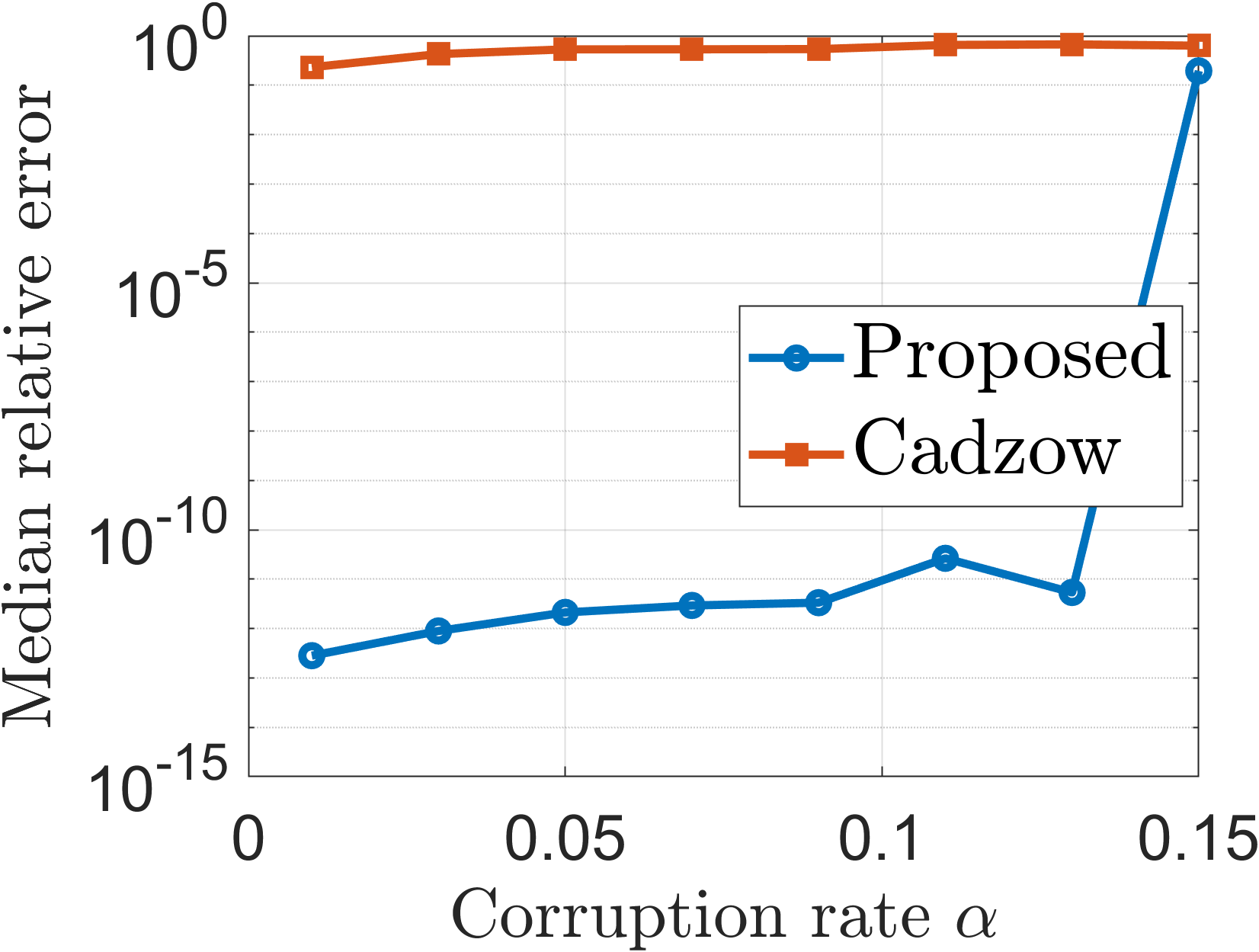}}\hfill
  \caption{Median relative error of spectral recovery versus corruption rate $\alpha$, computed over 15 independent trials per $\alpha$, with outlier amplification factors $c=1$ (\textbf{Left}) and $c=5$ (\textbf{Right}). Across a  range of \(\alpha\), the proposed method maintains near machine-precision accuracy, whereas the Cadzow baseline exhibits order-one error and is largely insensitive to decreasing \(\alpha\), reflecting its fragility to sparse snapshot corruptions.}
  \label{fig:re_vs_alpha_median}
\end{figure}

We fix \((d,m,J,L)=(21,3,7,300)\) and sweep the corruption rate
\(\alpha\in\{0.01,0.03,\ldots,0.15\}\).
For each \(\alpha\), we run 15 independent trials with \emph{sparse-in-time snapshot outliers only}
  and compute the spectral recovery relative error \(\mathrm{RE}\). 
We summarize variability across trials using a semilogy plot of the \emph{median} \(\mathrm{RE}\) versus \(\alpha\);
see \Cref{fig:re_vs_alpha_median}.
This experiment isolates the effect of increasing time-sparse corruptions and shows that explicitly modeling the
induced  Hankel outliers yields substantially improved robustness compared with purely Cadzow 
denoising-based baselines, for both mild outliers (\(c=1\), \textbf{Left}) and more severe corruptions
(\(c=5\), \textbf{Right}).

%  \begin{table}[h]
% \centering
% \caption{Signal-to-noise ratios (dB) for measurements and recovered spectra.}
% \label{tab:snr_results}
% \setlength{\tabcolsep}{7pt}
% \renewcommand{\arraystretch}{1.15}
% \begin{tabular}{rrrrr}
% \toprule
% $\mathrm{SNR}_{\text{out}}$ & $\mathrm{SNR}_{\text{Gauss}}$ & $\mathrm{SNR}_{\text{total}}$ & $\mathrm{SNR}_{\text{prop}}$ & $\mathrm{SNR}_{\text{cadz}}$ \\
% \midrule
% 7.7888  & 62.8783 &   7.7887&   35.8966  &  1.2921
% \\    
% 7.7888   & 82.7459  &   7.7889    &52.2721  &   1.2933\\
%     7.7888   &102.6463    & 7.7888    &76.9157   &  1.2940\\
%     7.7888  & 122.6463   &  7.7888  & 101.5926 &    1.2940\\
%     7.7888   &143.2371   &  7.7888   &110.4525   &  1.2940\\
%     7.7888  & 163.0564    & 7.7888   &148.6925   &  1.2940\\
% \bottomrule
% \end{tabular}
% \end{table}

\section{Conclusion and Future Work}
This paper proposes a novel robust spectral recovery model against time-sparse corruptions for dynamical sampling on a finite cyclic grid. A robust pipeline has been developed to solve the spectral recovery problem. The pipeline employs a sequence of robust Hankel recovery and completion algorithms to enable the stable Prony-style spectral recovery. A provable upper bound on the fraction of tolerable outliers has been established. The performance of the proposed methodology has been verified with empirical experiments. In particular, the proposed method tracks the true spectrum of the underlying convolution operator accurately and robustly, and outperforms the state-of-the-art constantly.

Future work will include a detailed theoretical analysis to bridge the spectrum of the convolution operator and the parameters of the lifted Hankel matrices (such as $\mu$ and $\kappa$). The analysis results will be used to motivate an improved design for the robust pipeline and algorithms. Extended empirical experiments, including real-world applications, will be evaluated. We will also extend the settings to high-dimensional dynamical sampling problems.

% We develop a robust spectral recovery framework for convolutional dynamical sampling with fixed uniform spatial subsampling under time-sparse snapshot corruptions, and we prove a noise-free low-rank guarantee for the associated Hankel liftings.
% These corruptions manifest as structured Hankel outliers, which we explicitly model to identify corrupted time indices and to perform robust Hankel reconstruction and completion across frequency channels.
% The resulting cleaned sequences enable stable Prony-style estimation of the spectrum of the underlying convolution operator, yielding accurate peak localization and significantly improved recovered-spectrum SNR.
% Across tested settings, the proposed method maintains high accuracy even at elevated corruption rates, whereas Cadzow-based baselines deteriorate markedly.

%\section*{Acknowledgment}

\newpage

\bibliographystyle{IEEEtran}
\bibliography{ref}

\end{document}